# Cloud Computing For Microfinances

**Suma. V**
Research and Industry Incubation Centre, Dayananda Sagar Institutions, India
sumavdsce@gmail.com

**Bhagavant Deshpande**
Assistant Professor, CIT ponnampet, India,
deshpacapricorn@gmail.com

**Vaidehi. M**
Research and Industry Incubation Centre, Dayananda Sagar Institutions, India
dm.vaidehi@gmail.com

**T.R. Gopalakrishnan Nair**
Endowed ARAMCO Chair, PMU,KSA,
Research and Industry Incubation Centre, Dayananda Sage Institutions, India
trgnair@ieee.org

## Abstract

Evolution of Science and Engineering has led to the growth of several commercial applications. The wide spread implementation of commercial based applications has in turn directed the emergence of advanced technologies such as cloud computing. India has well proven itself as a potential hub for advanced technologies including cloud based industrial market. Microfinance system has emerged out as a panacea to Indian economy since the population encompasses of people who come under poverty and below poverty index. However, one of the key challenges in successful operation of microfinance system in India has given rise to integration of financial services using sophisticated cloud computing model. This paper, therefore propose a fundamental cloud-based microfinance model in order to reduce high transaction risks involved during microfinance operations in an inexpensive and efficient manner.

## 1. INTRODUCTION

Since the advent of computers in 1945, there has been a tremendous growth in science and technology leading towards expansion of business in the industrial market. Management of massive amount of data due to the fast development of IT industries became one of the critical challenges to overcome in order to position themselves in the industrial atmosphere. Emergence of cloud as one of the upshots of technology has made a great impact on the business of several IT industries due to its varied advantages.

Figure 1 depicts the fundamental architecture of cloud computing. Accordingly, the architecture enables sharing of on-demand network resources to its varied clients with minimum management effort or service provider interactions.

Some of the widely used characteristics of cloud in industries include network-based services, scalability and assured quality of service to their customers which is comparatively inexpensive than the traditional network services and so on.

The cloud based industries facilities their users in terms of services such as Software-as-Service, Infrastructure-as-Service, and Platform-as-Service which can be either availed in isolated service mode or as an integrated service. The application programs that are offered through cloud computing is popularly known as Software as a service (SaaS) where wide spectrum of applications ranging from entertainment to business along with its essential resources are available to its clients. Additionally, cloud facilities the use of peripheral infrastructure in a highly flexible pattern which is well known as Infrastructure as a service (IaaS). However, unlike the classic client-server models, the centralization in cloud facilitates the service providers with absolute control over all the versions of browser-based applications, which are provided to clients. This in turn eliminates the dire need for clients to manage with version upgrades or licence management issues on client computing devices, which is generally known as Platform as Service (PaaS).

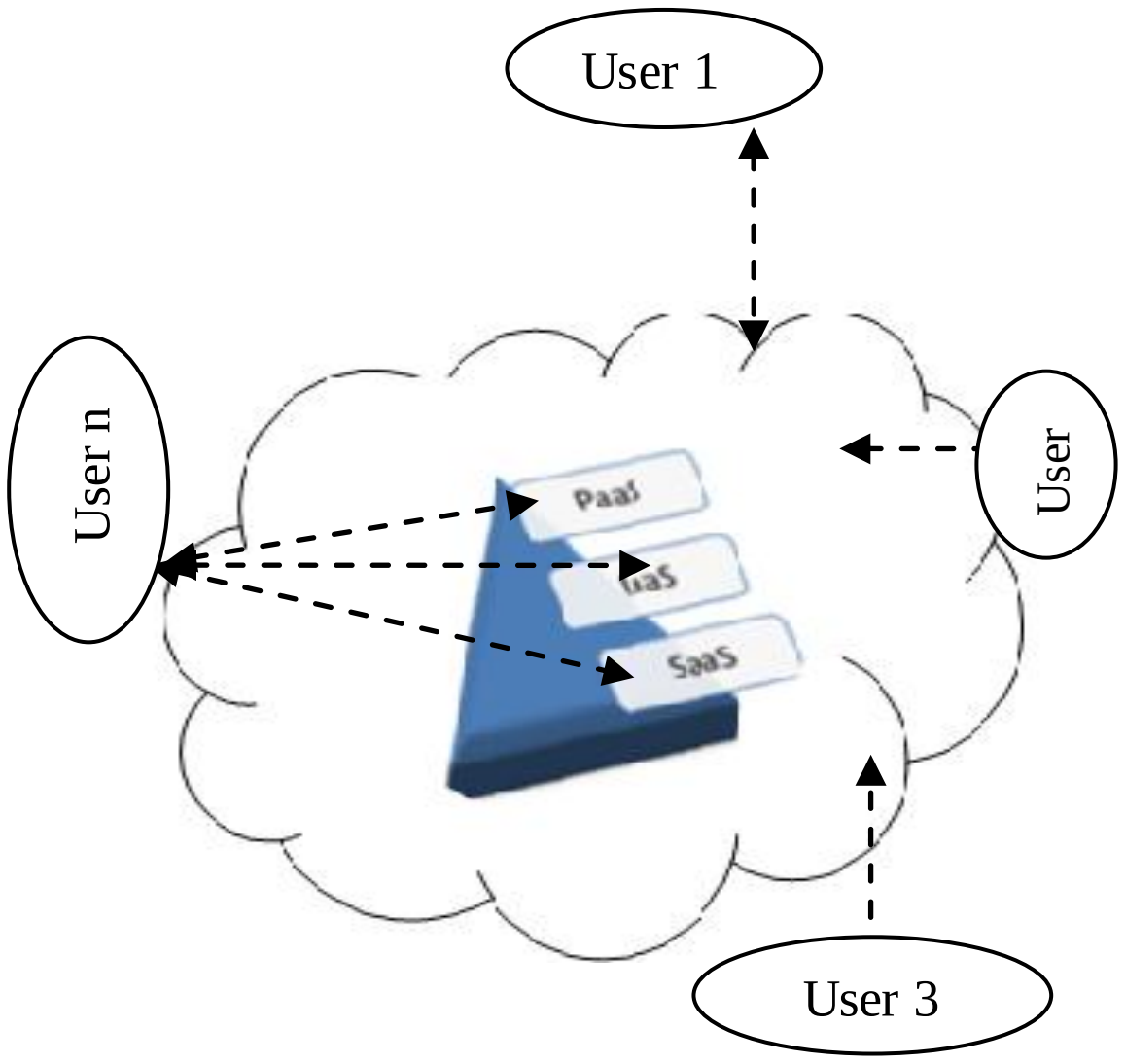

Fig. 1 Fundamental Cloud computing architecture

The wide spread implementation of cloud in software and hardware industries across the globe has made its foot also in Indian industrial market. The study made by author of [1] indicates progress of several Indian IT industries including academic institutions are moving towards cloud and thereby making India to be one of the globally potential cloud based market [1]. However, according to a survey conducted by the author in [2] states that there exists lack of awareness on cloud computing despite of the fact that many larger organizations are moving towards cloud technology. Nevertheless, the fast growth of Indian business in compliance with the emerging technologies, several segments such as microfinance are still in primitive state. This paper therefore presents a deeper analysis of microfinance sector of India in the cloud-based market.

## 2. Applications of cloud in Indian market

Cloud has made an impact in every domain of IT industries. India, being one of the leading IT hubs, has also oriented in the direction of cloud environment.

Authors of [2] states that small and medium enterprises (SME) of India contribute nearly 50% of Indian industrial economy. They strongly recommend the implementation of cloud computing in lieu of traditional ERP Systems.

However, authors [3] suggest the implementation of less expensive Enterprise Resource Planning (ERP) solutions to Indian industries using the latest advanced technologies such as cloud computing.

Authors of [4] have expressed the wide use of cloud in the academic domain in order to benefit the rural population in terms of knowledge and better education in an affordable cost package.

According to authors [5], the growth of software industries is one of the major national economy of India. They further suggest the implementation of cloud based software development in order to enhance the business.

Further, authors in [6] feels that cloud computing paradigm can be a promising technique to enhance the rural area development and henceforth the Indian economy.

However, India having considerably higher population of poor community has the highest potential for growth of microfinance. In support of the above fact, author of [7] suggest cloud computing to be implemented as a strategy to provide IT solutions in microfinance in order to eradicate most of the poverty residing in Indian population.

India is one of the leading and potential IT solution hubs and majority of Indian economy depends on the manner, in which economy can be generated from Indian population. Since, cloud computing is one of the proven technological advancement and microfinance being one of the challenges of Indian market, it is now possible to enhance the Indian financial system using cloud in microfinance as one of the betterment strategies.

## 3. Cloud based Microfinance

The root cause for growth of microfinance in India dates back to 1980 in order to aid the poor inhabitants. This being the main intent of establishment of microfinance, several non-governmental organizations has emerged out. The fast growth of microfinance organizations has resulted in massive data transactions. Author of [8] suggest a new paradigm for implementation of microfinance, which aim towards the reduction of transaction risks and costs. He further recommends the incorporation of efficient intermediate resources. India has furthermore proven itself in technological advancement and cloud computing is one of the latest growth of technology which is efficient, less expensive and improves the business of any organization. Hence, execution of microfinance services through cloud computing is deemed one of the potential contribution towards enhancement of Indian economy.

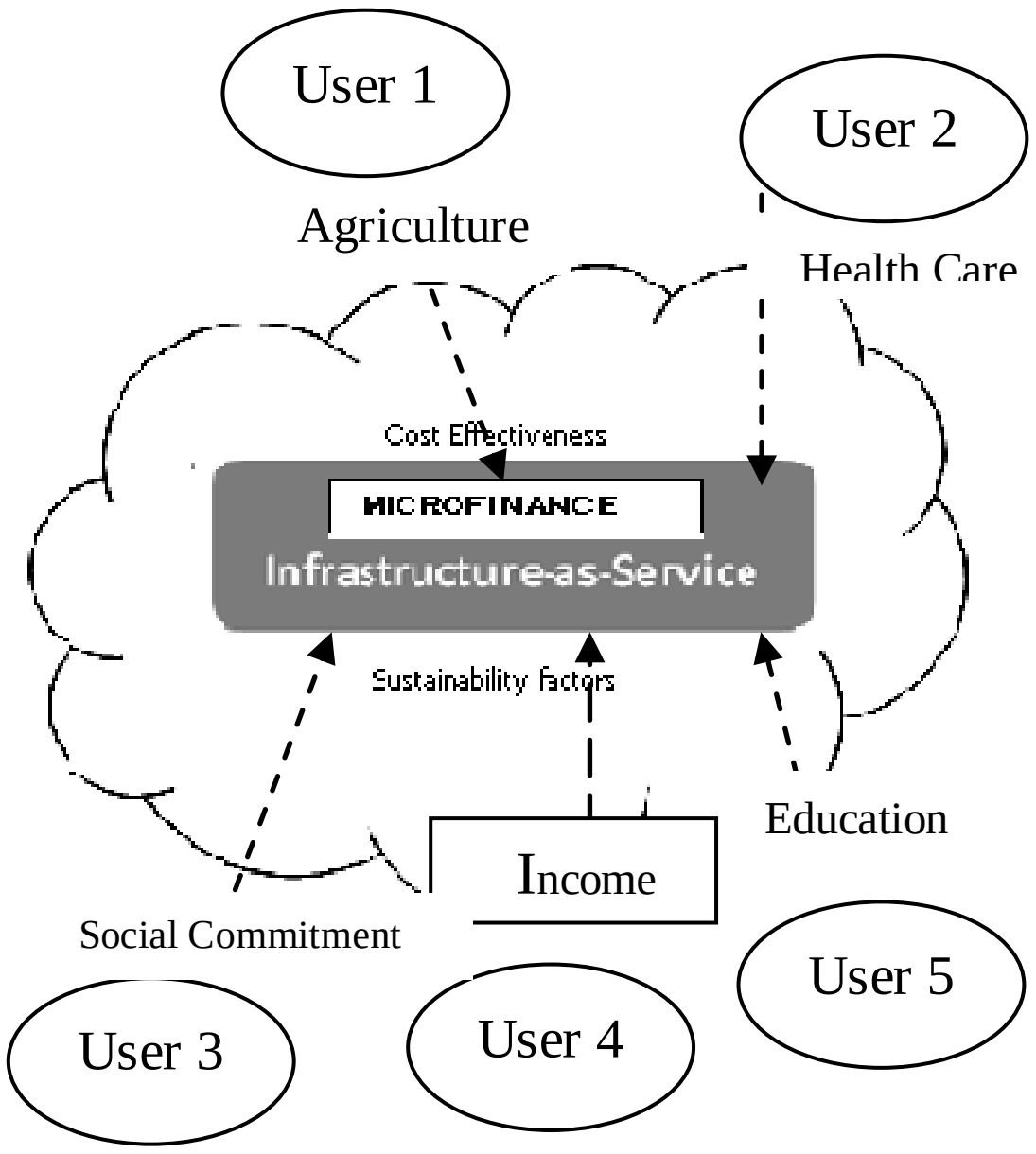

Fig 2. Cloud-based Microfinance model

Figure 2 depicts the fundamental cloud-based microfinance model. The figure infers that primary services offered by microfinance include financial assistance for agricultural activities, for health care, income, educational purposes and so on. These can be now be obtained in an inexpensive sophisticated cloud based environment which efficiently accelerates the financial transactions with lower risk, lesser human intervention and with improved performance to reach the entire population dwelling in poverty and below poverty line. This paper orients towards the introduction of cloud-based microfinance and our forthcoming papers explain the various issues concerned with cloud-based microfinance architecture and henceforth its remedial solutions.

## 4. Conclusion

The progression of knowledge in the domain of technology is one of the renowned contributions of Science and Engineering. This has given birth to several latest technological advancements such as cloud computing. India is one of the major contributors towards the growth of advanced technology to the global market. However, one of the lacunas of Indian economy is the population dwelling in poverty and below poverty baselines. Subsequently, the continual efforts made by several micro financial organizations have paved its way towards the improvement of Indian economy. The implementation of these micro financial services through the most modern cloud-computing technology is one of the promising strategies to improve the massive data management of Indian financial system. This paper proposes a cloud based microfinance model which is inexpensive and which functions with lower transaction risks in an efficient mode.